\documentclass[12pt,a4paper]{article}
\def\e{\begin{equation}}
	\def\f{\end{equation}}
\def\_#1{{\bf #1}}

\def\.{\cdot}

\def\=#1{\overline{\overline #1}}

\def\-#1{{\bf #1}}

\usepackage{amsmath}
\usepackage{lipsum}
\usepackage{graphicx,epstopdf}
\usepackage{epsfig}

\usepackage{amsfonts}
\usepackage{amssymb}
\usepackage{dcolumn}

\usepackage{soul}
\usepackage{color}

\usepackage{cite}

\begin{document}
\title{Reflectarrays and metasurface reflectors \\ as diffraction gratings}

\author{Fu Liu$^{1}$, Do-Hoon Kwon$^2$, Sergei A. Tretyakov$^3$\\
\\
\scriptsize{$^{1}$\it School of Electronic Science and Engineering, Xi'an Jiaotong University, Xi'an 710049, China}\\
\scriptsize{$^2$\it Department of Electrical and Computer Engineering, University of Massachusetts Amherst, Amherst, MA 01003 USA}\\
\scriptsize{$^3$\it Department of Electronics and Nanoengineering, Aalto University, P.O. Box 15500, FI-00076 Aalto, Finland}\\
}


\date{}
\maketitle

\begin{abstract}
Reconfigurable reflectors have a significant potential in future telecommunication systems, and approaches to the design and realization of full and tunable reflection control are now actively studied. Reflectarrays, being the classical approach to realization of scanning reflectors, are based on the phased-array theory (the so-called generalized reflection law) and the physical optics approximation of 
the reflection response. To overcome the limitations of the reflectarray technology, researchers actively study inhomogeneous metasurfaces, using the theory of diffraction gratings. In order to make these devices tunable and fully realize their  potential, it is necessary to unify the two approaches and study reconfigurable reflectors from a unified point of view. Here, we offer a basic tutorial on reflectarrays and reflecting metasufaces, explaining their common fundamental properties that stem from the diffraction theory. This tutorial is suitable for graduate and post-graduate students and hopefully will help to develop more deeper understanding of both phased arrays and diffraction gratings.  
\end{abstract}

\section{Introduction}

During recent few years, many research groups have been studying possible use of  reconfigurable intelligent surfaces (RIS) in future systems of wireless communications, e.g.  \cite{Liaskos,di2019smart,basar2019wireless,smart1,smart2,smart3,smart4,smart5,smart6,Marco_proc}. The main functionality of these metasurfaces is to reflect incident waves (coming from a specified direction or directions) into desired directions. Basically, this is the same function as is usually realized by reflectarray antennas. Most commonly, reflectarrays are used as flat or conformal equivalents of parabolic reflectors, while RIS are usually designed to reflect plane waves as plane waves propagating into  desired directions, but this is not a principal difference. Such metasurfaces are equivalent to focusing reflectors with an infinite focal distance. 

Recent research has shown that realizations of anomalously reflecting metasurfaces as phase-gradient reflectors (reactive impedance boundaries with a linearly-varying  phase of the local reflection coefficient) have a fundamentally limited efficiency, which degrades when the desired performance significantly deviates from that of uniform mirrors or retroreflectors \cite{Alu_2016,Elefth_2016,asadchy2016perfect,Diaz_From_2017,asadchy2017eliminating}. This degradation takes place because of excitation of parasitic propagating waves that scatter some part of the incident power into unwanted directions. Actually, similar effects are known also for reflectarrays, which function well only if the reflected rays do not have to be tilted much 
\cite{RA}. 
While for conventional applications of reflectarrays this problem can be tolerated, for the envisaged use of anomalous reflectors as reconfigurable intelligent surfaces, this limitation can significantly compromise their usability. Indeed, most usage scenarios assume that the reflected waves can be sent into any direction.

Recently, it was shown that advanced metasurfaces can control reflection theoretically perfectly, without any spurious scattering (except that caused by  manufacturing imperfections and  dissipation losses),  e.g.  \cite{Diaz_From_2017,asadchy2017eliminating,Radi_Metagrating_2018,Do-Hoon_2018,Diaz_Power_2019,Xuchen_multiple,budhu2021perfectly}. Different design approaches have been developed (we summarize and discuss them in Section~\ref{Sl}). Interestingly, all of them are based on the theory of diffraction gratings, and do not use the conventional design methods and topologies that have been developed for phased arrays and reflectarray antennas. 

It appears that for proper understanding and further development of devices for full and efficient control of wave reflection it is necessary to analyze the basic principle of inhomogeneous reflectors, looking at both subwavelength-structured metasurfaces and reflectarrays formed by repeating antenna elements at half-wavelength intervals from a unified point of view. While these two techniques are different,  there are fundamental similarities  of metasurface reflectors and reflectarrays: both can be considered as diffraction gratings. 

In this basic tutorial paper, we explain the fundamental principles behind any device that creates plane waves propagating in a certain direction from the point of view of the theory of diffraction gratings.
In the last section, we summarize and classify the currently known methods to design and realize anomalous reflectors and discuss current research challenges.

\section{Active arrays}

The main challenge in the design of reflectarrays and reflecting metasurfaces is to ensure that incident waves excite such currents on the reflector that create the desired reflected fields. However, first of all one needs to know what current distribution is necessary to realize. To this end, we will  discuss 
active arrays~\cite{mailloux2017}, 
assuming that we can fix any desirable distribution of radiating current over a planar surface. Our goal is to determine what current distribution we should set in order to create the desired set of propagating waves. Here, it is enough to consider sheets of electric surface currents. Although these sheets create waves on both sides of the sheet (in actual implementations, a ground plane or a complementary sheet of magnetic surface current are used), we will be able to properly determine the needed surface profile of the current distribution over the reflector plane even using this simple model.  

For simplicity, we will consider infinite arrays, and our desired reflected modes will be plane waves. For infinite arrays, the most common design goal is to ensure that enough far from the array surface, where all the evanescent fields of the array elements can be neglected, there is only one plane wave propagating in the desired direction. For finite arrays, this design goal is equivalent to ensuring that the radiation pattern has only one main beam pointing in the desired direction, without any grating lobes. In this sense, the conclusions made for infinite arrays will hold also for finite arrays.

\subsection{The  period of the radiating current distribution for a given radiation direction}

Let us suppose that the reflected field that we want to create in the far zone is a set of propagating plane waves. We will assume that the fields of this desired set of plane waves vary along the planar radiating surface (the coordinate $x$) as a  \emph{periodical} function. This means that the tangential wavenumbers of all radiated harmonics are in rational relations. We make this assumption for simplicity and in view of the very important special case when we want to launch only one  obliquely propagating plane wave with the transverse wavenumber $k_t=k_0\sin\theta$, where $k_0$ is the free-space wavenumber. The $x$-dependence of this field, function $e^{-jk_0x\sin\theta}$, is a periodical function with the period ${\lambda/{\sin\theta}}$, where $\lambda={2\pi/k_0}$ is the wavelength.
The case of launching aperiodically distributed fields can be in principle treated as a limiting case of the infinite period. Later, we will also discuss possibilities to launch a single plane wave with aperiodical current distributions.

It is clear that the radiating current distribution should be in phase synchronism with all the  desired free-space modes which we want to launch. Assuming that the radiating current distribution is periodic, let us expand this periodic current distribution into spatial Fourier series. Denoting the period of the radiating current distribution as $D$, the radiating current will have Fourier components with tangential wavenumbers 
\e k_{tn}= {2\pi n\over D},\qquad n=0,\pm 1,\pm2,\dots\label{current}\f 
In general, we should select the period $D$ so that the tangential wavenumbers of all plane waves which we want to create will be found among this set of numbers (for some values of index $n$).

Let us consider the special case of launching only \emph{one plane wave} at a certain angle $\theta$. In this case, it is enough to properly set only one Fourier harmonic of the current.
The tangential wavenumber of the desired plane wave is 
\e k_t=k_0\sin \theta={2\pi\over \lambda} \sin \theta   \label{field}\f
We need to select $D$ so that at least one harmonic of the current distribution is in  phase with the desired radiated wave. Comparing Eqs.~\eqref{current} and \eqref{field}, the condition reads
$n/D=\sin\theta/\lambda$, that is, $D={n\lambda/\sin\theta}$.
It is reasonable to choose as small $n$ as possible (that is, as small $D$ as possible) to minimize the number of ``open channels'' or diffraction maxima (the directions where the array can radiate), because we want to send the energy only in one direction. $n=0$ is not a valid solution, since in that case $k_t=0$, so we select $n= 1$, which gives
\e 
D={\lambda\over \sin\theta}\label{D}\f 
In this case, the period of the radiating current is equal to the period of variations of the fields in the plane wave that we want to create. Obviously, this is an expected result.

Very importantly, we note that the array period $D\ge \lambda$ for any angle $\theta$. The limit $D\rightarrow\lambda$ corresponds to $\theta\rightarrow \pi/2$, that is, to the end-fire array. For small angles (radiation directions close to the normal), the period is very large. 
As a specific numerical example, we will consider arrays that create a single plane wave in $\theta=70^\circ$ direction. 
For this example angle $D \approx 1.0642 \, \lambda$.

We have already noted that it is desirable to select as small $D$ as possible to minimize the number of propagating harmonics. More specifically, for a given value of the period $D$ all harmonics that  have the tangential wavenumber $k_{tn}$ satisfying the inequality 
\e |k_{tn}|={2\pi\over D}|n|< k_0={2\pi\over \lambda}\f
are propagating modes. This corresponds to \e |n|<{D\over \lambda}\f
In our example case of radiation into $70^\circ$ direction, we have $|n|<1.0642$, that is, 
if we set \emph{any} periodical current on the antenna (using active sources, recall that in this section we discuss active arrays), with this value of the period, the radiation (in the far zone) can go only to $0$, $70^\circ$, and $-70^\circ$ directions. All higher-order harmonics are evanescent. 
Note, however, that for scanning into other directions the situation can be very different, since we may need $D$ which is rather large as compared with the wavelength. 

We see that from the diffraction theory point of view, any periodical antenna array that radiates in any direction except the normal direction is a \emph{diffraction grating}, because its period $D$ is larger than the wavelength, and it radiates into an $n\ne 0$ spatial harmonic.

\subsection{The optimal current distribution}

Because the period of the radiating current distribution is larger than the wavelength, the array can create more than one propagating plane wave.
The next task is to set such current distribution that the waves into all the unwanted directions would have zero amplitude.

Let us denote the current distribution along the antenna plane as $J(x)$. As before, $x$ is the coordinate along the array plane. In our example of excitation of a single plane wave, the tangential fields of the desired plane-wave mode vary as $e^{-jk_0x\sin\theta}$.  As discussed above, the current distribution on the antenna should be a periodical function with the same period as this plane wave, $D=\lambda/\sin\theta$. But what specific function with this period we should select? 
An appropriate current distribution can be found by considering the excitation integrals (integrals of the current distribution and the complex-conjugate of the desired radiated field distribution). It is enough to integrate over one period.  

Continuing discussing a specific example of launching a single plane wave along $70^\circ$ direction, when only three plane waves can propagate, the current distribution should be such that
\e  
\int_{0}^{D}  J(x) e^{jk_0x\sin\theta} dx \quad \rightarrow \quad \mbox{maximum}\label{max}\f
\e 
\int_{0}^{D}  J(x) e^{-jk_0x\sin\theta} dx =0,\qquad \int_{0}^{D}  J(x) dx =0 \label{1}\f
These are the coupling integrals with all the three propagating modes that can be excited with the selected period of the antenna. Importantly, we do not impose any restrictions on the amplitudes of evanescent fields in the vicinity of the array, as in thought applications the goal is to create the desired field in the far zone. Note, however, that in reflectarray and metasurface realizations evanescent modes need to be controlled in order to optimize far-zone fields, and we will briefly discuss it in Section~\ref{Sl}.

Clearly, there are many possible solutions for the current profile satisfying Eqs.~\eqref{max}--\eqref{1}. The simplest and obvious one is the current of a constant amplitude and a linear phase gradient. Selecting 
$J(x)=J_0 e^{-jk_0x\sin\theta}$ we maximize the integral in \eqref{max} because the product of the two exponentials is just unity. The first integral in \eqref{1} is zero because of double variations over the period, which is why we do not radiate into the $-70^\circ$ direction. Obviously, the second integral in \eqref{1} is also zero, ensuring no radiation in the normal direction. 
That is why currents with this ideal linear phase profile  radiate a perfect single plane wave.

Let us consider another case when the tilt angle is small, say $\theta=5^\circ$. We use the same simple theory and first select the period such that we couple to plane waves along this direction. Now the appropriate period is
\e
D={\lambda\over \sin\theta} \approx 11.4737 \,\lambda\f
The plane-wave Fourier harmonics will propagate when 
$k_t= 2 \pi |n| /D < k_0$, that is, when 
$|n|< D /\lambda = 11.47$.
Thus, our desired direction corresponds to 
$n=1$, but there are 22 more directions
(11 on the left, 10 on the right side
from the normal, and the normal direction)
where the waves can also propagate. 
The current distribution should be such that
\e  
\int_{0}^{D}  J(x) e^{jk_0x\sin\theta} dx \quad \rightarrow \quad \mbox{maximum}\label{maxlong}\f
\e
\int_{0}^{D}  J(x) e^{jnk_0x\sin\theta} dx=0,\quad n=0,-1,\pm2,\pm3,\dots\label{long}
\f

The current distribution with the ideal linear phase gradient clearly satisfies all these relations. For approximate (step-wise, for example) settings finding solutions which would radiate only in one single direction is a non-trivial task. On the other hand, for small angles the required linear phase variation is smooth, and coupling to other propagating harmonics is weak. Thus, reasonably smooth discretized distributions of the linear phase function create nearly perfect single-wave fields in the far zone.

\subsection{Realization with small radiating elements}\label{sec:radielem}

Next, let us discuss the principle of diffraction grating (metagrating) \cite{Radi_Metagrating_2018,Popov_Controlling_2018,Epstein_Unveiling_2017,rabinovich_ieeejap2018,rabinovich_prb2019} realization. Instead of a continuous current distribution, let us excite the desired plane wave with just a few small radiating elements placed in each period. To understand how it works, we assume that the current distribution over the antenna plane is a set of a few delta-function sources. That is, we consider an array of small isotropic radiators. Let us try to reach the goal using only two such small active radiating elements per period. We position them at two arbitrary selected points of each period, say $x=0$ and $D/2$.

The current distribution function becomes 
$J(x) =A \delta(x)+ B \delta(x-D/2)$. 
Here $A$ and $B$ are the complex amplitudes of the two point sources (line sources in the 2D scenario).
Substituting this sum of two delta functions into \eqref{1} we come to a system of two linear equations for two unknown amplitudes $A$ and $B$, which, however, gives only a trivial solution.  
This means that we need at least three elements. Let us assume 
\e J(x) =A \delta(x)+ B \delta(x-D/3)
+C \delta(x-2D/3)\label{delta-pattern}\f
and substituting it to Eq.~\eqref{max}--\eqref{1}, we get
\e B=-A(1+e^{j2\pi/3})=Ae^{-j2\pi/3},\qquad C=-(A+B)=Ae^{j2\pi/3} \label{eq:elemBC}\f
to ensure no radiation into the normal and the $-70^\circ$ directions. 
We see that the three line sources have the same amplitude but with a linear phase drop in this case of equal spacing (total $2\pi$ phase drop over one period). By substituting these values of $C$ and $B$ into \eqref{max}, we find that the result is $3A$ (nonzero), which means that in the far zone there is a perfect plane wave propagating into the desired direction. The positions of the small elements can be considered as varying parameters, adding degrees of freedom in design. Here we note a recent paper \cite{Hansen_WM} that studied excitation of plane waves by arrays of line currents. In that paper, excitation conditions are imposed also on evanescent modes, resulting in conclusion that only continuous current distributions with a linear phase gradient can launch a single plane wave.

\begin{figure}[b!]
	\centering
	\includegraphics[width=0.5\textwidth]{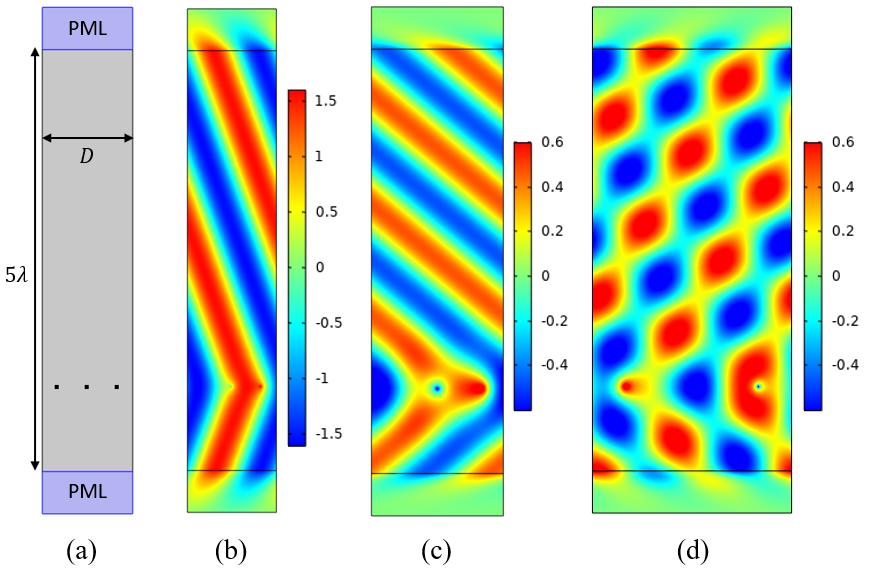}
	\caption{Radiation from  three discrete currents, with the current distribution function given by Eqs.~(\ref{delta-pattern})--\eqref{eq:elemBC}. Note that the supercell shown here has a lateral shift along the $x$ axis. (a) Schematic of the simulation, where the currents are flowing in the out-of-screen direction. (b-d) The scattered electric field pattern for angles $\theta=70^\circ$, $40^\circ$, and $25^\circ$, respectively.}
	\label{fig:3pointsrc}
\end{figure}

The above results are verified with numerical simulations, as presented in Fig.~\ref{fig:3pointsrc}. When the three line sources (2D point sources shown in the figure) are assigned with the current distribution accordingly to Eqs.~(\ref{delta-pattern})--\eqref{eq:elemBC} with $A=1$~mA, we indeed see that a perfect plane wave is generated in the desired direction when $\theta>30^\circ$, as shown in Figs.~\ref{fig:3pointsrc}(b) and (c). However, when $\theta\leq30^\circ$, the scattered field is not a single plane wave, as shown in Fig.~\ref{fig:3pointsrc}(d). The working angles $\theta>30^\circ$ correspond to $|n|<D/\lambda<2$, which means that three point sources are enough for generating perfect plane waves when only the modes $|n|<2$ are propagating modes.

For the case of many propagating modes the design complicates. For the above example of radiation into the direction of $\theta=5^\circ$ we need to satisfy  a system of 23 equations. 
However, we can use a simple approach based on Eqs.~(\ref{delta-pattern})--\eqref{eq:elemBC}.
Similarly, for $N$ discrete sources which are evenly distributed over $D$, we can assume that the current distribution follows
\e
	J(N,x)=\sum_{m=0}^{N-1} Ae^{-j2\pi m/N} \delta(x-Dm/N)\label{eq:JxNp}
\f
i.e., with the same amplitude and a linear phase drop of $2\pi$ over $D$. Then, the integration
\e
	\int_{0}^{D} J(N,x) e^{jnk_0 x\sin\theta } dx \label{eq:intJ}
\f
shows where there is a plane wave going to port $n~(n\in\mathbb{Z})$: zero value indicates that there is no power sent into port $n$, while a non-zero value indicates that there is a plane wave emitted to port $n$. In Table~\ref{tab:intJ} we summarize the integration results for different $N$ and $n$. As we can see, when $N=3$, the integration is non-zero for ports $n=-5,-2,1$, and $4$, which means that if these ports correspond to propagating modes, then there will be plane waves propagating to those ports. This is consistent with the results in Fig.~\ref{fig:3pointsrc}(d), where for $\theta=25^\circ$, there are plane waves propagating in ports $1$ and $-2$ (verified by simulations).

\begin{table}[t!]
	\centering
	\caption{The integration results for different number of discrete sources $N$ and port $n$.}
	\label{tab:intJ}
	\begin{tabular}{c|ccccccccccccc}
		\;	& \multicolumn{13}{c}{$n$}\\
		$N$	&-6	&-5	&-4	&-3	&-2	&-1	&0	&1	&2	&3	&4	&5	&6\\ \hline
		2	&0	&$2A$&0	&$2A$&0	&$2A$&0	&$2A$&0	&$2A$&0	&$2A$&0\\
		3	&0	&$3A$&0	&0	&$3A$&0	&0	&$3A$&0	&0	&$3A$&0	&0\\
		4	&0	&0	&0	&$4A$&0	&0	&0	&$4A$&0	&0	&0	&$4A$&0\\
		5	&0	&0	&$5A$&0	&0	&0	&0	&$5A$&0	&0	&0	&0	&$5A$\\
		6	&0	&$6A$&0	&0	&0	&0	&0	&$6A$&0	&0	&0	&0	&0\\
	\end{tabular}
\end{table}

\begin{figure}[b!]
	\centering
	\includegraphics[width=0.5\textwidth]{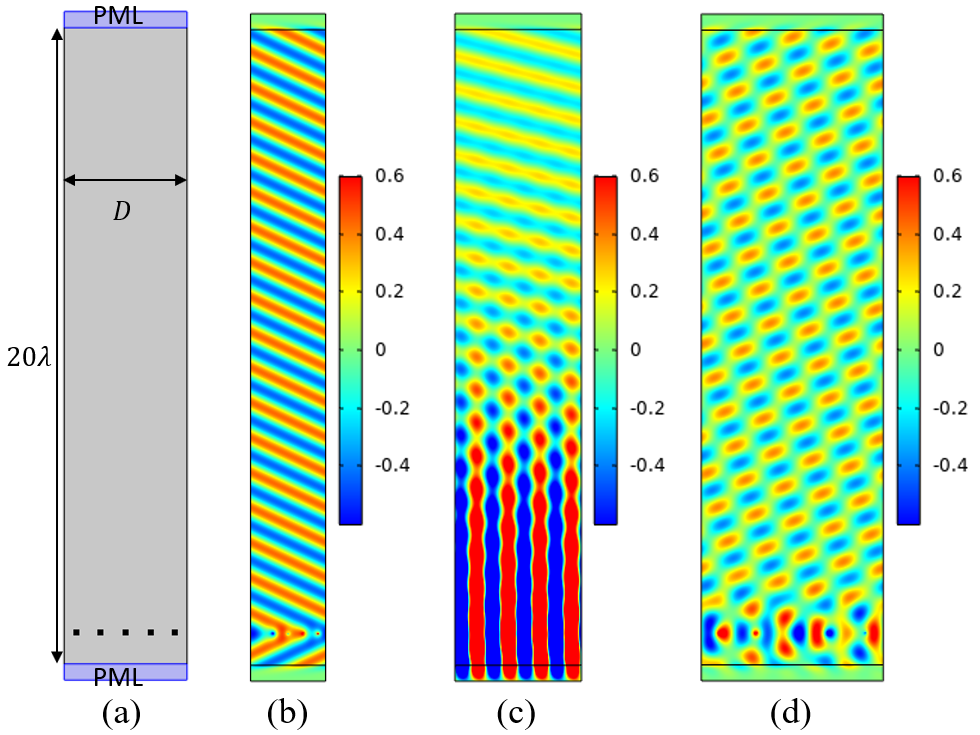}
	\caption{Radiation from  $N=5$ discrete currents in one supercell period $D$ (with lateral shift along $x$). (a) Schematic of the simulation. (b-d) The scattered electric field pattern for angles $\theta=25^\circ$, $14.5^\circ$, and $10^\circ$, respectively.}
	\label{fig:5pointsrc}
\end{figure}

To validate these results, the fields of $N=5$ discrete sources are also simulated and the simulation setup and results are shown in Fig.~\ref{fig:5pointsrc}. As we can see, when $\theta>14.48^\circ$ [Figs.~\ref{fig:5pointsrc}(b,c)], which corresponds to $|n|<D/\lambda<4$, there is only one plane wave propagating to port $n=1$ ($\theta$ direction). However, when $\theta=10^\circ$, where the ports $n=\pm4$ are also open, there is also a plane wave propagating in port $n=-4$, resulting in  interference of plane waves, as shown in Fig.~\ref{fig:5pointsrc}(d).

From Table~\ref{tab:intJ} we also see that as the number of discrete sources $N$ increases, more ports correspond to evanescent modes and only the port $n=1$ corresponds to a propagating wave. Indeed, as  $N$ increases, the discrete current distribution is closer to the analytical continuous current distribution with a linear phase gradient.

\begin{figure}[b!]
	\centering
	\includegraphics[width=0.65\textwidth]{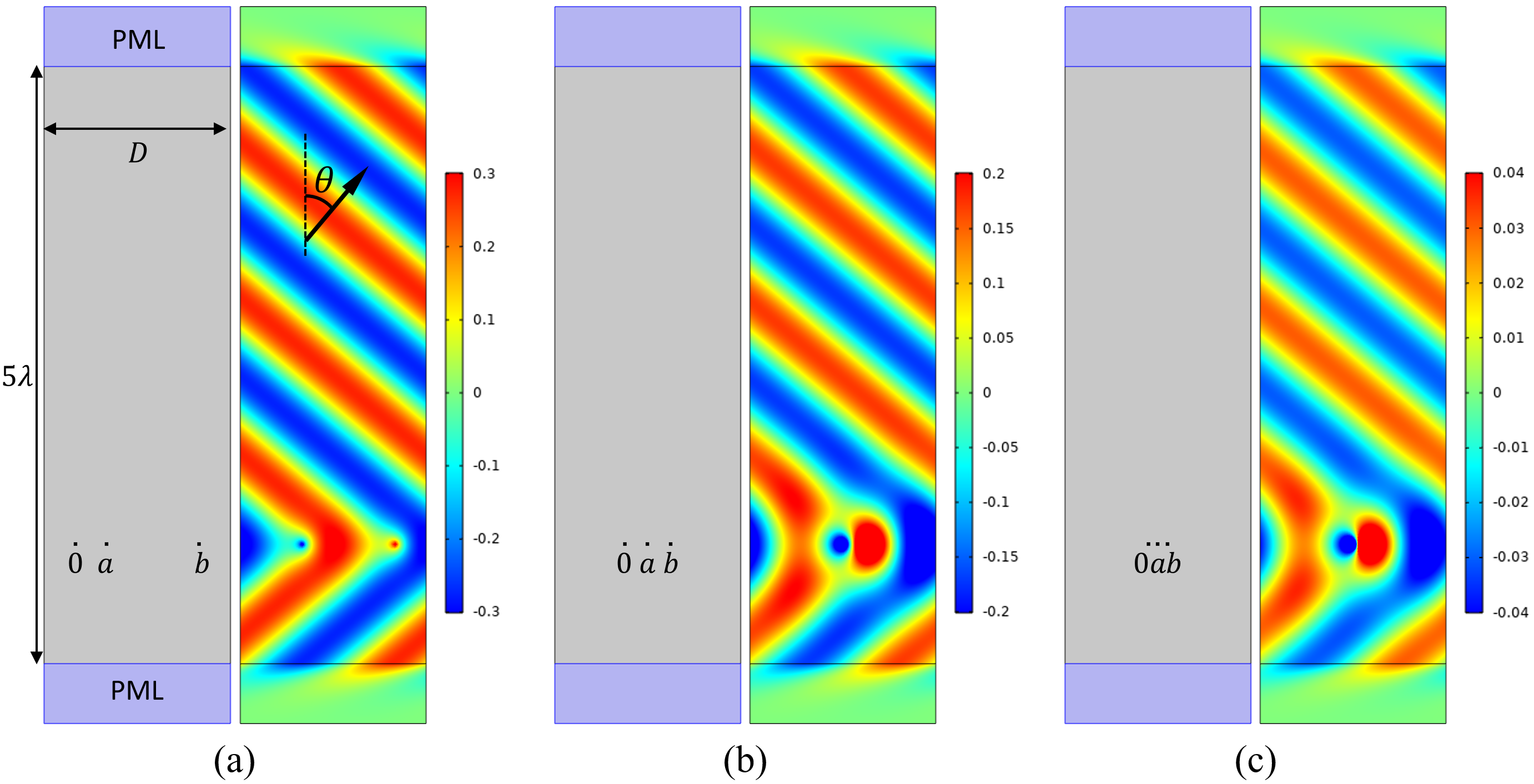}
	\caption{Radiation from  $N=3$ discrete current sources that are arbitrarily positioned in one period (with lateral shift along $x$). Each of the sub-figures shows the schematic of the simulation and the radiated electric field pattern for  $\theta=40^\circ$. The assumed values are: (a) $a=D/6,~b=2D/3$; (b) $b=2a=D/4$; (c) $b=2a=D/10$. The complex source amplitudes $B$ and $C$ can be obtained from Eq.~\eqref{eq:BCcoeff} with $A=1$~mA.}
	\label{fig:3pointsrc_arbitrary}
\end{figure}

It is important to note that, in this design approach, the radiating elements do not have to be equally spaced. Indeed, instead of \eqref{delta-pattern}, let us assume 
\e J(x) =A \delta(x)+ B \delta(x-a) +C \delta(x-b)\label{delta-pattern_nu}\f
where $a,b<D$. By substituting it to Eq.~\eqref{1} to eliminate the wave radiation in the normal  and $-\theta$ directions ($n=0$ and $n=-1$), 
we find that the solution of $B$ and $C$ in terms of $A$ are
\e
B=A\dfrac{-1-e^{j2\pi a/D}+e^{j2\pi b/D}+e^{j2\pi(a-b)/D}}{2-2\cos(2\pi(b-a)/D)},~
C=A\dfrac{-1+e^{j2\pi a/D}-e^{j2\pi b/D}+e^{j2\pi(b-a)/D}}{2-2\cos(2\pi(b-a)/D)}\label{eq:BCcoeff}
\f
It is clear that, when the radiating elements are not equally spaced, the required phase distribution does not follow the linear phase gradient law. In addition, the amplitudes of the currents are also changed. For example, when we set $a=D/6$ and $b=2D/3$, as shown in Fig.~\ref{fig:3pointsrc_arbitrary}(a), the required complex amplitudes of the second and third sources read $B=(-\frac{3}{4}-\frac{\sqrt{3}}{4}j)A,~C=(-\frac{1}{4}+\frac{\sqrt{3}}{4}j)A$, and with this setting we can still get the desired single-wave radiation, as illustrated in Fig.~\ref{fig:3pointsrc_arbitrary}(a).

If we further set $b=2a$ and make $a$ much smaller than $D$, so that the three sources are clustered,  we find the complex amplitudes as
\e B=A(-1-e^{j2\pi a/D}),\quad C=Ae^{j2\pi a/D} \f
This means that the phase is quickly varying among the three sources, i.e., the first and third sources have small phases $0$ and $2\pi a/D$ ($a\ll D$) while the middle source has a phase that is  close to $\pi$, and the amplitude of the middle source is close to 2. As examples, in Fig.~\ref{fig:3pointsrc_arbitrary}(b) and (c), we perform two sets of simulations with $b=2a=D/4$ and $b=2a=D/10$, respectively. The results show that these clustered sources can still generate the desired radiation without any unwanted scatterings. We note that it is straightforward to analyze similarly the case of small tilt angles with more discrete point sources. Obviously, realization of such fast variations of the reflection phase in reflectarrays is challenging, and it is preferred to use equal spacing.

\subsection{Realization with radiating segments}


In actual realizations of antenna arrays, for example, using patch antennas, the radiating currents are not point sources, rather, they are  small radiating elements. To analyze this case, we consider discretization of the current distribution into segments. In each segment, the current is uniform, while for different segments currents have different phases, still following  the linear phase profile in the step-wise fashion. In this case, we can write the current distribution as
\e
    J(N,x)=Ae^{j\Phi(N,x)}\label{eq:Jxs}
\f
where $N$ is the number of segments in each period $D$, $A$ is the complex amplitude of the currents, and the phase function
\e
    \Phi(N,x)=-2\pi\frac{{\rm Floor}(xN/D)}{N}-\delta\phi  \label{eq:phis}
\f
corresponds to the discretized phase profile, with $\delta\phi$ being a phase-shift factor. 

\begin{table}[t!]
	\centering
	\caption{The absolute value of integrals  Eq.~\eqref{eq:intJ} for the current distribution  Eq.~\eqref{eq:Jxs} with port number $n$ and segment number $N$.}
	\label{tab:intJd}
	\begin{tabular}{c|ccccccccccccc}
		\;	& \multicolumn{13}{c}{$n$}\\
		$N$	&-6	&-5	&-4	&-3	&-2	&-1	&0	&1	&2	&3	&4	&5	&6\\ \hline
		2	&0	&$0.13A$&0	&$0.21A$&0	&$0.64A$&0	&$0.64A$&0	&$0.21A$&0	&$0.13A$&0\\
		3	&0	&$0.16A$&0	&0	&$0.42A$&0	&0	&$0.83A$&0	&0	&$0.21A$&0	&0\\
		4	&0	&0	&0	&$0.3A$&0	&0	&0	&$0.91A$&0	&0	&0	&$0.18A$&0\\
		5	&0	&0	&$0.24A$&0	&0	&0	&0	&$0.94A$&0	&0	&0	&0	&$0.16A$\\
		6	&0	&$0.2A$&0	&0	&0	&0	&0	&$0.96A$&0	&0	&0	&0	&0\\
	\end{tabular}
\end{table}

With such segmented current distribution, we can  calculate the integral in Eq.~\eqref{eq:intJ}, and the value will indicate if there is a plane wave propagating to each port $n$. Similarly to Table~\ref{tab:intJ}, Table~\ref{tab:intJd} shows the absolute values of the integral for $\delta\phi=0$. As we can see, the results show the same pattern as  Table~\ref{tab:intJ}, i.e., the positions of zero and non-zero values are the same, while there is only some difference in the amplitudes (and also phases) of the integrals. As a result, all the conclusions in Subsection~\ref{sec:radielem} are valid also in this case, and the simulation results will give very similar field patterns shown in Figs.~\ref{fig:3pointsrc} and~\ref{fig:5pointsrc} (verified with COMSOL simulations). We further note that the change of the phase shift factor $\delta\phi$ in Eq.~\eqref{eq:phis} does not change the integration results at all, meaning that the phase can be arbitrarily shifted.

Actually,  we can replace the point sources or uniform-current sections by some arbitrary array elements, say of a square shape or a cos-shape and check if the radiation remains perfect for the same number of elements per period.
Instead of \eqref{delta-pattern}, we  assume the following current pattern:
\e J(x) =A F(x)+ B F(x-D/3)
+C F(x-2D/3), \label{eq:J_segment_1}\f
where $F(x)$ is the current pattern at each array element and $A,B,C$ are the complex amplitudes of currents at the array elements. 
For example, we can assume a square shape or a sin-shape
\e F(x)=\begin{cases} 1 & \text{if} \quad 0<x<w \\
0 & \text{otherwise}
\end{cases},\qquad
F(x)=\begin{cases} \sin(\pi x/w) & \text{if} \quad 0<x<w \\
	0 & \text{otherwise}
\end{cases} \label{eq:cos_shape}
\f
where $w$ is the width of the array element which is assumed to be smaller than $D/3$.  Then, following the same approach as before, i.e., solving equations \eqref{1} by substituting the current pattern, we find that the conditions for radiation only to the desired angle $(n=1)$ is when the complex amplitudes are
\e B=Ae^{-j2\pi/3},~~C=Ae^{j2\pi/3} \f
which are independent from the element width $w$. We note that this solution is exactly the same as the solution \eqref{eq:elemBC} for point-source realization in Section~\ref{sec:radielem}, clearly indicating that the point sources and uniform-current sections can be replaced by arbitrary radiating elements. In fact, this conclusion holds true even when the segments are not equally spaced over the period. For example, if we assume the current pattern as
\e J(x) =A F(x)+ B F(x-a) +C F(x-b),\label{eq:J_segment_2}\f
similarly to  Eq.~(\ref{delta-pattern_nu}) for the point source, the required complex amplitudes are also found to be given by Eq.~\eqref{eq:BCcoeff}, which is also independent of the segment width $w$.
\begin{figure}[b!]
	\centering
	\includegraphics[width=0.75\textwidth]{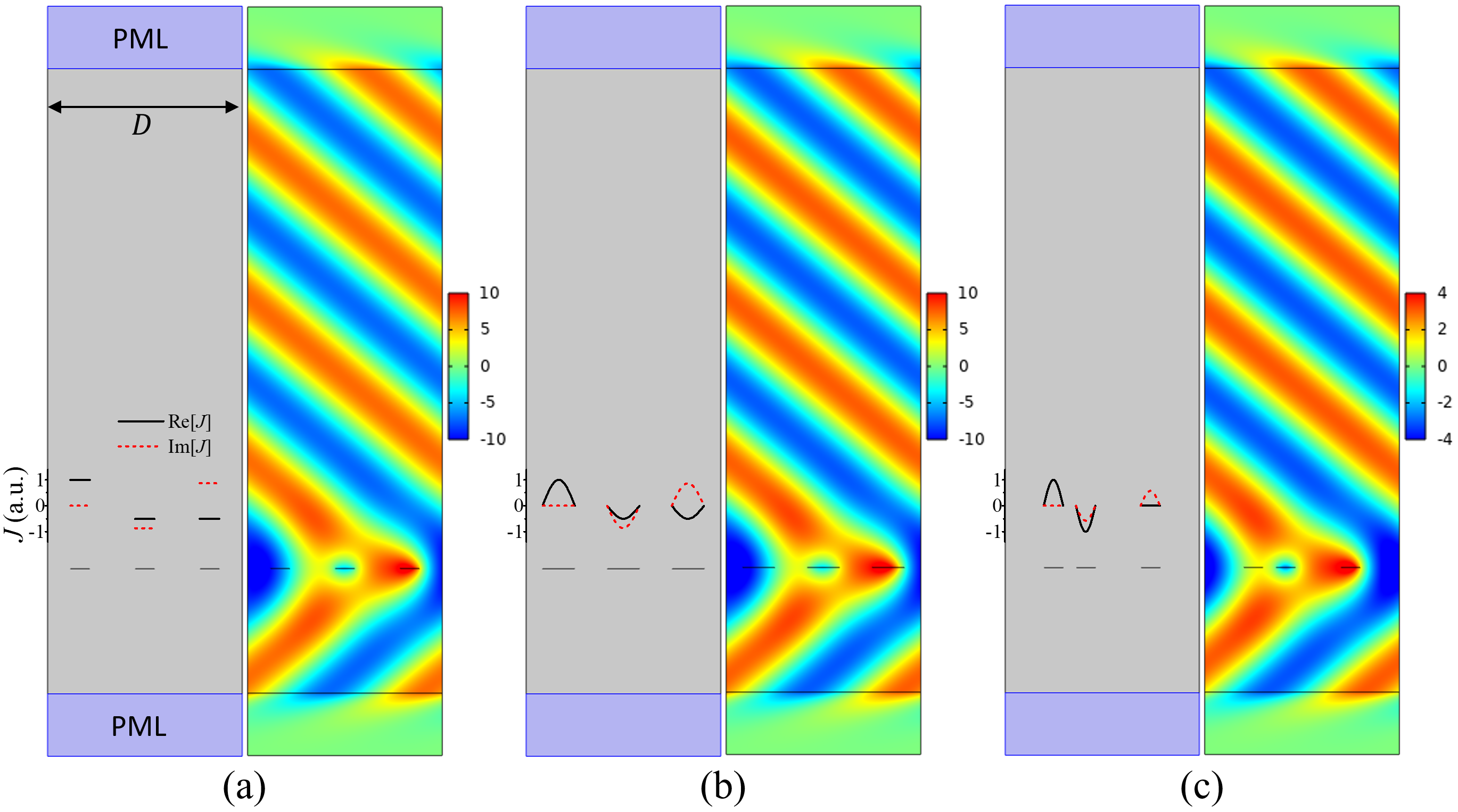}
	\caption{Radiation from  $N=3$ radiating segments per period (with lateral shift along $x$). Each of the sub-figures shows the schematic of the simulation and the electric field pattern for  $\theta=40^\circ$. The insets show the complex current patterns. The configurations of them are: (a) equally spaced segments with $w=D/10$ and square-shape current profile; (b) equally spaced segments with $w=D/6$ and sine-shape current profile; (c) arbitrarily spaced segments with $a=D/6,~b=D/2,~w=D/10$ and cosine-shape current profile. $B$ and $C$ are obtained from Eq.~\eqref{eq:BCcoeff} and $A=0.1$~A/m is assumed for all simulations.}
	\label{fig:3segmentsrc}
\end{figure}

To demonstrate these realizations, we perform three sets of simulations, presented in Fig.~\ref{fig:3segmentsrc}. All the three configurations, i.e., equally-spaced radiating segments with square-shape current pattern in Fig.~\ref{fig:3segmentsrc}(a), equally-spaced radiating segments with cosine-shaped current pattern in Fig.~\ref{fig:3segmentsrc}(b), and arbitrarily-spaced radiating segments with cosine-shape current pattern in Fig.~\ref{fig:3segmentsrc}(c), radiate only to the desired angle $\theta=40^\circ$ that corresponds to $n=1$.

The assumption of a real-valued function $F(x)$ corresponds to the case of the use of small resonant 
antennas as array elements. This is the case when the current distribution over a unit element is a standing wave, and the current phase is uniform over the unit cell. Also, arrays of slots in metal sheets can be modelled this way. However, it is possible to assume that the current distribution is a complex-valued function. The conclusions will not change. 

It is obvious that these models assume that the current distribution over each array element is fixed and independent from the excitation of this element and all the other elements in the array.
In the case of reflectarrays or metasurfaces, this is equivalent to the assumption that the distribution of near fields in the vicinity of the array does not depend on the incidence angle.

\subsection{Diffraction-grating approach \emph{versus} phased-array approach}

In the above theory, we started from stating that current distributions that launch a single plane wave into a desired direction should be periodical functions, with the period related to the period of the plane wave that is radiated. The reason for that assumption is that the excitation current and the field of the mode that we excite must be in phase synchronism. The other reason is that fixing the period as defined by the desired reflection angle or angles, one makes sure that the reflector can create a plane wave exactly in the desired direction. The design goal is then to properly distribute the reflected power among the allowed ``open channels'', minimizing scattering into all directions except the desired one. From the practical point of view, design of periodical structures reduces to the design of only one period, which is a great simplification.

However, the current distribution does not have to be a periodical function. Conditions \eqref{max}--\eqref{1} or \eqref{maxlong}--\eqref{long} can be considered as conditions on the spatial Fourier transforms of the current distributions, if we assume that $J(x)$ is not necessarily periodical and extend the integration over the whole $x$-axis. 
Actually, Floquet-periodical functions 
of the form
\e J(x)= \sum_{m=-\infty}^\infty A_m e^{-j\left(k_0\sin\theta +{2\pi m\over d}\right)x}\label{Floq_form}
\f
can satisfy the required conditions for launching only one plane wave in the direction given by the angle $\theta$.
Formula \eqref{Floq_form} corresponds to an ideal current profile $e^{-jk_0\sin\theta x}$ 
that is modulated by an arbitrary  periodical function $F(x)$ with the period $d$: $J(x)=F(x)e^{-jk_0\sin\theta x}$. Expanding $F(x)$ into Fourier series, we arrive at \eqref{Floq_form}.

Really, the term corresponding to $m=0$ is the ideal current profile $e^{-jk_0\sin(\theta) x}$ (uniform amplitude, linear phase gradient) corresponding to a current distribution that launches a single plane wave in the direction defined by the angle $\theta$. Indeed, this function is obviously orthogonal to all other plane waves as eigenmodes of free space, because 
\e 
\int_{-\infty}^\infty e^{-jk_0\sin(\theta) x}e^{jk_0\sin(\theta') x}dx=0  \quad \mbox{for all}  \ \theta'\neq \theta
\label{ort}
\f 
Thus, if we ensure that also all other members of the series \eqref{Floq_form} have this property, currents of the form \eqref{Floq_form} will excite only one single propagating plane wave. 
To ensure this property, we demand that
\e \left|k_0\sin\theta +{2\pi m\over d}\right|>k_0\f
for all $m\neq 0$. In this case, the tangential components of the wave vectors of all harmonics with $m\neq 0$ are larger than $k_0\sin\theta'$ for all $\theta'$, and the orthogonality condition \eqref{ort} is satisfied.
This relation is convenient to rewrite as
\e  \left|\sin\theta +{\lambda \over d}m\right|>1\f
For $\theta\rightarrow 0$ the condition is satisfied if $d<\lambda$ (the most ``dangerous'' is, obviously,  the term $m=-1$). For $\theta\rightarrow \pi/2$, the condition is true if $d<\lambda/2$. That is why for scanning phased arrays the period $d$ is conventionally chosen to be equal to $\lambda/2$: this choice ensures that for any scan angle no parasitic diffraction lobes will be created. 

The case when the current distribution is not a periodical function is equivalent to the limit of a periodical function with an infinite period. As we see,  although there are infinitely many allowed plane-wave
propagation directions when the period approaches infinity, it is possible
to limit the number of \emph{propagating} plane-wave direction
to one by proper phasing of the currents at the phased-array elements.

If the radiation direction is fixed and no scan is required, the discretization period $d$ can be set for  that specific angle as
\e d<\lambda/(1+\sin\theta)\label{eq:dcond}\f
For instance, for our example of $\theta=70^\circ$, we have $d<0.5155 \lambda$. This is also verified from the results of Figs.~\ref{fig:3pointsrc} and \ref{fig:5pointsrc}: 
\begin{itemize}
	\item For $N=3$, the critical angle is $\theta=30^\circ$, which gives the critical period $D=\lambda/\sin\theta=2\lambda$ or the critical distance $d=D/3=2\lambda/3$, which is consistent with Eq.~\eqref{eq:dcond}.
	\item For $N=5$, the critical angle is $\theta=14.48^\circ$, which gives the critical period $D=\lambda/\sin\theta=4\lambda$. Thus, the  critical distance $d=D/5=4\lambda/5$, which is consistent with Eq.~\eqref{eq:dcond}.
\end{itemize}

The periodical ``modulation function'' $F(x)$ can be a complex-valued function. An important special case is the function of the form $f(x')e^{jk_0\sin\theta x'}$, defined in the region $-d/2<x'<d/2$. Periodically repeating this modulation, we have in the period number $M$ (where $x=Md+x'$) the current in the form
\e J(x)=F(x)e^{-jk_0\sin\theta x}=f(x')e^{jk_0\sin\theta(x-Md)}
e^{-jk_0\sin\theta x}=f(x')e^{-jMdk_0\sin\theta}\label{far}\f
This is the case of the
phased-array design approach. The structure is formed as an array of unit cells of size $d$ (the antenna array elements) each of which has the same current pattern $f(x')$, but the phase of these unit-cell current elements linearly varies from cell to cell. The continuous linear phase profile is replaced by a discrete sum. The most common assumption is the case when function $f(x')$ is real-valued, which corresponds, for example, to arrays of resonant dipoles, resonant patches, or arrays of horn antennas. 

We see that the phased-array approach  is quite different from the conventional approach to the design of diffraction gratings and metasurfaces for reflection control. Here, we start from fixing the unit-cell size $d$, usually at $d=\lambda/2$. Assuming the same amplitude distribution over each unit cell (usually the resonant mode of an antenna array element), we adjust the phases of the unit cells to create constructive interference in the desired direction.



\begin{figure}[b!]
	\centering
	\includegraphics[width=0.55\textwidth]{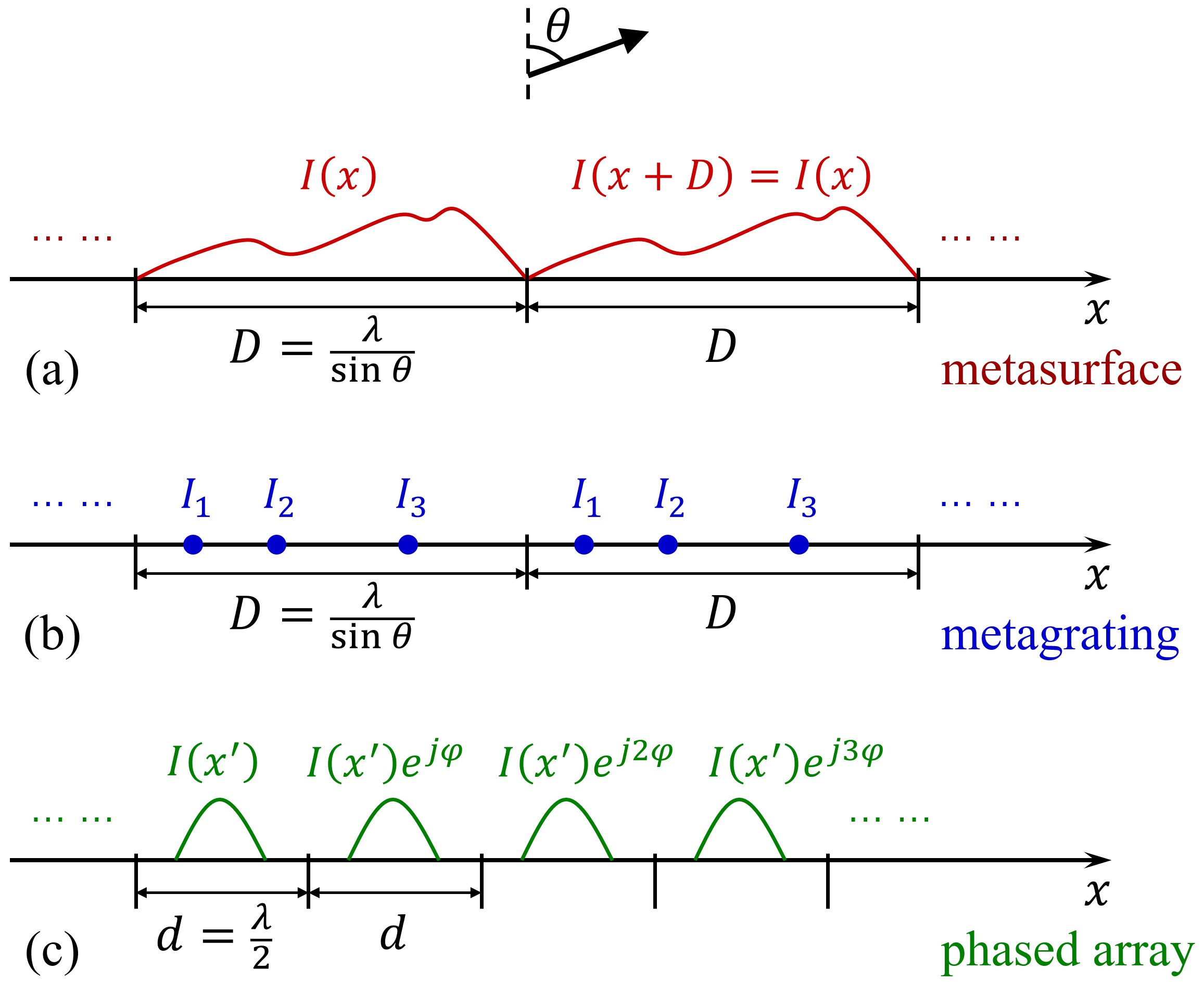}
	\caption{(a) Metasurface design approach (effectively continuous periodical current distribution); (b) Metagrating design approach (a few small scatterers in each period); (c) Phased-array design approach (in general, aperiodical current distribution). }
	\label{dp}
\end{figure}

Figure~\ref{dp} illustrates these two approaches. The top two pictures correspond to the diffraction-grating design approach to designing anomalous reflectors.   In this case, we select a proper period of the current distribution $D=\lambda/\sin\theta$ and find a suitable current distribution (the solution is not unique). For the case of a discretized current [point elements, Fig.~\ref{dp}(b)] to launch the wave to $70^\circ$ where we have three propagating Floquet modes, we need  three discrete radiating sources in each period $D=\lambda/\sin\theta$.

In Fig.~\ref{dp}(c) that illustrated the phased-array approach,  we have an array of ``patches'', all of the same shape and size and spaced by $\lambda/2$. We assume that the amplitude pattern is the same for all units (a cos-shape is shown) and that the phase is uniform over each unit cell. Then we feed this array with a linearly varying phase source, so that the phases vary, from unit to unit, as 
\begin{equation} \Phi(x)=-j{2\pi\over \lambda} \sin(70^\circ) x=
-j{2\pi\over \lambda} \sin(70^\circ) n{\lambda\over 2}
\end{equation}
as shown in the picture.
Importantly, we note that this current distribution is, in general, \emph{not a periodical function.} The periodicity condition (assuming the uniform amplitude) reads
\e n\pi \sin\theta=2\pi N\f
where $N$ is the number of unit cell that will have the same phase as the cell number 0. Obviously, this condition can be satisfied only if $\sin\theta$ is a rational number.

\begin{figure}[b!]
	\centering
	\includegraphics[width=0.9\textwidth]{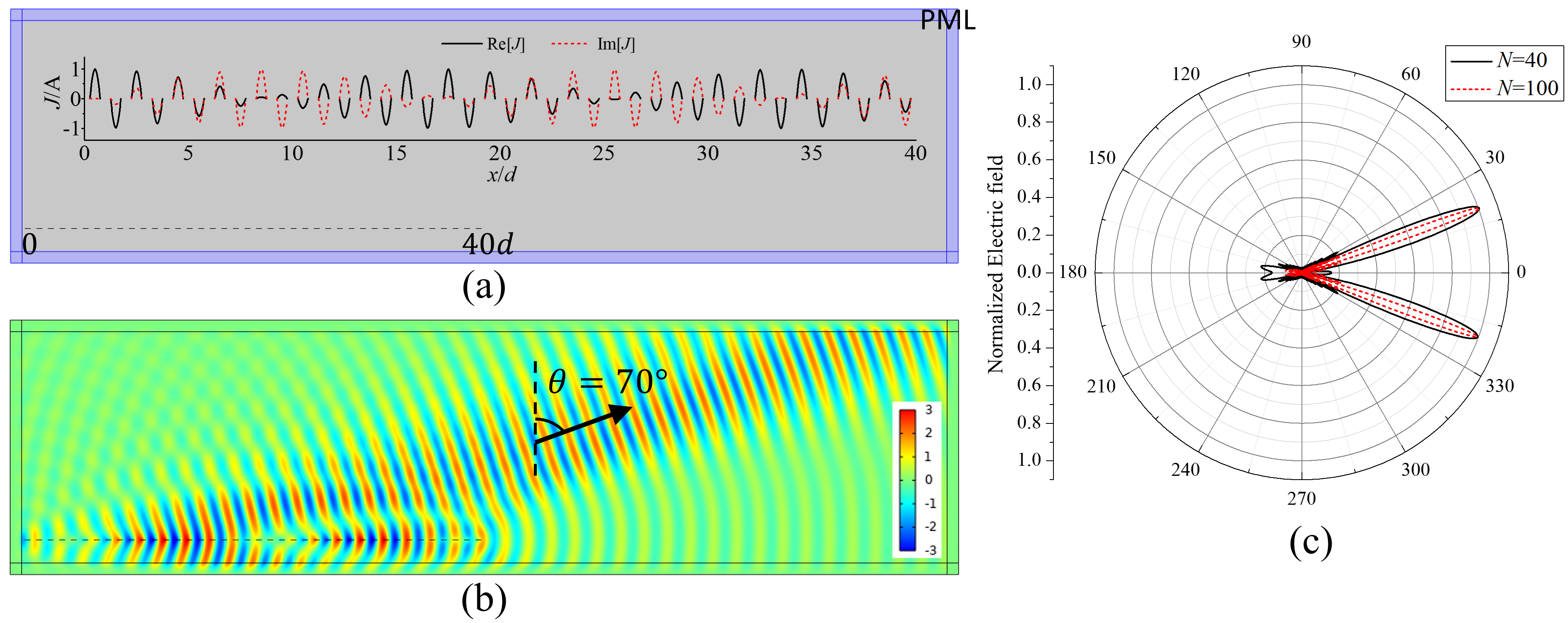}
	\caption{Radiation from a finite phased array consisting of $N=40$ ``patches'' with width $w=\lambda/4$ and spacing $d=\lambda/2$. (a) Setup of the simulation. The inset shows the current distribution. (b) Simulated electric field pattern from the finite phased array. (c) Normalized far field (electric field) emitted from the finite phased array. The black solid (red dashed) line corresponds to the phased array consisting of $N=40$ and $N=100$ ``patches''. }
	\label{fig:phasedarray}
\end{figure}

Figure~\ref{fig:phasedarray} shows an example of the active phased array for generating radiation toward $\theta=70^\circ$. Due to the non-periodicity, we assume a finite number of ``patches'' with the width $w=\lambda/4$ and spacing $d=\lambda/2$ to form a finite phased array. The current of this finite phased array is set to follow Eq.~(\ref{far}) with $f(x')$ being a cos-shape, as shown in the inset of Fig.~\ref{fig:phasedarray}(a). It indeed radiates in the desired direction, as shown by the electric field pattern in Fig.~\ref{fig:phasedarray}(b) and the normalized far field (black solid curve) in Fig.~\ref{fig:phasedarray}(c). Due to the finiteness of the array, there are side lobes, but no other main beams. With increasing the number of ``patches'', the pattern tends to a delta function. This is illustrated by the red dashed curve in Fig.~\ref{fig:phasedarray}(c) which is for the case with $N=100$.

If we choose the element spacing in the
phased array equal to $D/3$, the phase array becomes periodical, and it is equivalent to
the diffraction grating picture with the
element current changed from a spatial delta to a cosine distribution.
Hence, we note that  for active arrays the difference
between diffraction grating and phased array approaches
is of practical nature in essence. In diffraction gratings, 
a periodic physical configuration stipulated by the incidence and
refraction angles is chosen so that the design domain
is reduced to a single supercell. In phased arrays,  the cell size (i.e., the element
spacing) is determined \emph{a priori} 
(usually, $\lambda/2$) based on the required
beam scanning capability. The functionality of creating one
propagating wave is performed by element excitations
following a spatially linear phase profile.

The fundamental difference between the two design approaches is in realizing scanning capability. For diffraction gratings, the period of the array should be adjusted for each desired reflection angle, which in practice requires enough small, subwavelength unit cells. For phased arrays, the period is fixed, which in principle allows the use of $\lambda/2$-sized unit cells. In this case, the required current distribution over the plane is, in general, aperiodic. This issue does not create practical problems for active phased arrays, which are all of transmitting type. However, realization of the required linear phase profile for wide-angle scans by adjusting reactive bulk loads of passive reflectarray antenna elements is rather difficult. 


Let us stress that there is a very important simplifying assumption in the root of the phased-array antenna theory. Namely, it is  assumed that the current \emph{distribution} over one array element does not depend on the currents on the other elements. The current density on the first element of one period can be written as 
$I_0 f(x)$, where $f(x)=0$ at $x<0$ and $x>d$, and $I_0$ is the current amplitude fixed by external sources. Only under this assumption, the current amplitude of the element number $M$ is $I_Mf(x-Md)$, where $f$ is \emph{the same} function. 
This assumption is an approximation. It holds very well for active resonant antenna elements when the current distribution is approximately fixed by the resonant mode or for arrays of horn antennas where the current (the aperture field) is very close to that of the fundamental mode of the horn. We stress that the current amplitudes $I_M$ depend on external voltages applied to \emph{all} elements, but knowing the impedance matrix of the array we can always find such voltages exciting active elements that realize the desired distribution of current amplitudes over the antenna plane.

\section{Reflectarrays and metamirrors}
\label{Sl}

In contrast to active antenna arrays where each antenna element is fed by an external, controllable source, currents on elements of reflectarrays and metasurfaces are excited by incident waves (and by fields scattered by all other array elements). The usual design aim is to  synthesize the induced current distribution $J(x)$  to create a constant-amplitude, linear phase profile. However, unlike in the active phased array case, we do not have full control over the magnitude and phase of $J(x)$ using a passive surface subject to illumination by incident waves. 

Let us first consider the design approach based on the diffraction-grating theory. That is, we fix the array period as defined by the incidence angle and the desired reflection direction and design a surface to perform the required operation. In the following discussion, we will consider the same example functionality as in the examples above: reflection of a normally incident plane wave into a plane wave into an arbitrary direction. 
We can consider two cases:

\begin{enumerate}
    \item If we specify the scattered field as a single plane wave, for example, at the $70^\circ$ tilt, and calculate the required surface impedance $Z_{s}$ of the reflector, an alternately active/lossy surface results \cite{asadchy2016perfect,Estakhri_Wavefront_2016}. This is not desirable or even possible from the  realization perspective.

\item If we specify the periodically varying surface reactance to give a unit-magnitude, linear phase gradient  local reflection coefficient profile in the locally periodic approximation, passivity is guaranteed. However, the reflected field ends up including parasitic propagating waves. In our example, also into $-70^\circ$ and $0^\circ$ directions, e.g.  \cite{Diaz_From_2017}. Other (and more than two) parasitic scattering directions open up for other combinations of  the incidence and anomalous reflection angles.  

\end{enumerate}
Neither case is satisfactory.
Researchers work on finding other solutions that use only passive (with as small dissipation as possible) array elements. The only possibility is to allow excitation of other field modes in addition to the only desired plane wave propagating into the desired direction. If all these other modes do not propagate into far zone (if all are evanescent Floquet harmonics), the perfect performance of the device is not compromised.    

There can be several ways to approach this task:

\begin{enumerate}
    \item 
Model the reflector as a boundary defined by its surface (input) impedance. Allowing excitation of surface-bound modes, design the surface impedance  so that only the desired reflected propagating mode exists, enforcing ${\rm Re}[Z_s(x)]=0$ as an additional condition. Finding $Z_s(x)$ for creating only one propagating plane wave is simple, but guaranteeing passivity is not straightforward. This approach has been used, for example, in  \cite{DoHoon_Lossless_2018} and \cite{budhu2021perfectly}. Importantly, actual realizations of the optimized surface impedance still requires the use of locally periodical approximation, possibly leading to performance degradation. For this reason, the version developed in \cite{Marco_proc} imposes an additional condition of slow variation of the surface impedance that can make the approximate design of surface topology more accurate. 

\item Start from a certain topology (usually a patterned metal sheet on a grounded dielectric substrate) and optimize the evanescent fields so that the effective sheet impedance of the thin patterned metal layer is lossless at every point. This is the approach used in \cite{Xuchen_multiple} and  \cite{Do-Hoon_planar_2021}
for scalar metasurfaces and
in \cite{yepes_apl2021} for tensor metasurfaces.
In this approach, the impedance model is used only for a single patterned sheet, not for the whole metasurface structure. Also in this method the locally periodical approximation is used, but only for the design of the reactive sheet, instead of the whole metasurface body.

\item
Considering arrays of thin metal strips on a grounded substrate and using the periodic Green's function for a grounded dielectric substrate, find the loading impedances of strips analytically. Translate the required loading element into a gap capacitor, as is done, e.g., in \cite{rabinovich_ieeejap2018,rabinovich_prb2019}. This approach is a variation of the metagrating design method. This way it is possible to  treat anomalous transmission as well as reflection~\cite{rabinovich_ieeejap2020}.

\item Design the current distribution so that in the far zone only the desired plane wave is created by direct optimization of the metasurface structure (not relying on sheet or surface impedance boundary conditions and not requiring that the input impedance is reactive at any surface, except the ground plane). Passivity is guaranteed by microscopically considering passive inclusions, but creating only one plane wave is not as easy as in 1). This approach has been used in \cite{Diaz_From_2017} and \cite{wong_prx2018}.

\item Find a surface where the input impedance is purely reactive for the desired set of waves and design the local surface reactance $X_s(x)$ \cite{Tereshin,Diaz_Power_2019}.  This is the power flow-conformal solution, which does not need optimizations of spatial dispersion properties of the reflector.

\end{enumerate}

Let us next discuss the design approach based on the phased-array theory. That is, we fix the unit-cell size, usually to  $\lambda/2$. We position geometrically-identical passive antenna elements into each cell and find such loads connected to the elements that the induced current distribution has the required linear phase profile (with a periodical modulation, as in \eqref{far}) . 

Under the simplifying assumption that the current distribution over each antenna element is the same for any incident field distribution and any loads connected to all the elements, we can use the impedance-matrix method to find the load impedances that realize the desired phase distribution. Unfortunately, we run into the same problem as with the diffraction-grating design: these loads have active-lossy behavior  \cite{Tereshin}. Thus, one needs to impose an additional constraint on the load impedances (zero real parts) and design some optimization procedure for finding reactive loads that approximate the required amplitude and current distribution in the best possible way. Basically, we end up with the same problem of engineering near fields so that the power is properly channeled from ``virtually lossy'' to ``virtually active'' antenna elements (engineering spatial dispersion). It is not clear if just one control element per $\lambda/2$ will be enough to reach the goal, and we may need to use subwavelength elements, similarly to the metasurface scenario. 

The main difficulty in the phased-array (reflectarray) approach is that the array of antennas  (including the loads) is no longer
a periodical structure. This makes the number of ports for
connecting load impedances infinite, and all the loads are in general different. In contrast, the periodical metasurface approach requires design and optimization of only one period, and usually only a few parameters need to be optimized. We stress that this it not a problem in a conventional practical 
phased array antenna in either transmission or receiving mode because it corresponds
to a transmissive device. A transmissive device does not have 
the interference problem between the incident and 
scattered waves, which is the reason for the active-lossy nature of the required surface impedance of an ideally performing anomalously reflecting boundary.

\section{Conclusion}

As we discussed in the introduction, in the recent literature on metasurfaces for control of reflected waves, there have been many publications on the suggested use of reconfigurable anomalous reflectors for engineering and optimization of propagation environment. Vast majority of these works is based on the locally-periodical approximation of the array response. That is, for calculations of the reflected fields locally defined reflection coefficient is used, and it is assumed that it can be somehow controlled by changing the parameters of the unit cell at each position. This is the conventional reflectarray antenna design assumption. It is known that this local design gives acceptable performance for moderate tilt angles (moderate deviations from the reflection law for uniform mirrors). 
However, for envisaged  applications it is required that beams can be directed into any desired direction, which requires more advanced designs. Majority of the current research  in this field uses variations of periodical metasurfaces (diffraction-grating design methods). Here, impressive results have been achieved in design and realization of anomalous reflectors for fixed sets of incidence and reflection angles. The main problem here is the realization of electrical reconfigurability. Because in this design approach the period of the array is  defined by the incidence and reflection angle, it should be possible to change the array \emph{period} by adjusting tunable components of small unit cells. 

On the other hand, the phased-array approach uses a set of periodically arranged antennas, and the geometrical period is fixed for all scan angles. However, the distribution of controllable loads is not periodical in this case, which requires global optimization of the whole array for each scan angle. Moreover,  it is not known if optimization of loads of a conventional $\lambda/2$-spaced array can lead to acceptable performance for 
required extremely wide scan-angle range. Approaches to optimization of reactive loads of reflectarray antennas can be found e.g. in \cite{Tereshin,El_arxiv},   but it  appears that more research on this design approach is needed.

\section*{Acknowledgements}
This work was supported in part by the European Integrated Training Network METAWIRELESS and the US Army Research Office grant W911NF-19-2-0244.

\bibliographystyle{IEEEtran}
\bibliography{IEEEabrv,references}

\begin{thebibliography}{10}
\providecommand{\url}[1]{#1}
\csname url@samestyle\endcsname
\providecommand{\newblock}{\relax}
\providecommand{\bibinfo}[2]{#2}
\providecommand{\BIBentrySTDinterwordspacing}{\spaceskip=0pt\relax}
\providecommand{\BIBentryALTinterwordstretchfactor}{4}
\providecommand{\BIBentryALTinterwordspacing}{\spaceskip=\fontdimen2\font plus
\BIBentryALTinterwordstretchfactor\fontdimen3\font minus
  \fontdimen4\font\relax}
\providecommand{\BIBforeignlanguage}[2]{{%
\expandafter\ifx\csname l@#1\endcsname\relax
\typeout{** WARNING: IEEEtran.bst: No hyphenation pattern has been}%
\typeout{** loaded for the language `#1'. Using the pattern for}%
\typeout{** the default language instead.}%
\else
\language=\csname l@#1\endcsname
\fi
#2}}
\providecommand{\BIBdecl}{\relax}
\BIBdecl

\bibitem{Liaskos}
C.~Liaskos, S.~Nie, A.~Tsioliaridou, A.~Pitsillides, S.~Ioannidis, and
  I.~Akyildiz, ``A new wireless communication paradigm through
  software-controlled metasurfaces,'' \emph{IEEE Commun. Mag.}, vol.~56, no.~9,
  pp. 162--169, 2018.

\bibitem{di2019smart}
M.~{Di Renzo}, M.~Debbah, D.-T. Phan-Huy, A.~Zappone, M.-S. Alouini, C.~Yuen,
  V.~Sciancalepore, G.~C. Alexandropoulos, J.~Hoydis, H.~Gacanin \emph{et~al.},
  ``Smart radio environments empowered by reconfigurable ai meta-surfaces: An
  idea whose time has come,'' \emph{EURASIP Journal on Wireless Communications
  and Networking}, vol. 2019, no.~1, pp. 1--20, 2019.

\bibitem{basar2019wireless}
E.~Basar, M.~{Di Renzo}, J.~De~Rosny, M.~Debbah, M.-S. Alouini, and R.~Zhang,
  ``Wireless communications through reconfigurable intelligent surfaces,''
  \emph{IEEE Access}, vol.~7, pp. 116\,753--116\,773, 2019.

\bibitem{smart1}
W.~Tang, J.~Dai, M.~Chen, X.~Li, Q.~Cheng, S.~Jin, K.~Wong, and T.~J. Cui,
  ``Subject editor spotlight on programmable metasurfaces: The future of
  wireless?'' \emph{IET Electron. Lett.}, vol.~55, no.~7, pp. 360--361, 2019.

\bibitem{smart2}
Q.~Wu and R.~Zhang, ``Towards smart and reconfigurable environment: Intelligent
  reflecting surface aided wireless network,'' \emph{IEEE Commun. Mag.},
  vol.~58, no.~1, pp. 106--112, 2020.

\bibitem{smart3}
{\"O}.~{\"O}zdogan, E.~Bj{\"o}rnson, and E.~G. Larsson, ``Intelligent
  reflecting surfaces: Physics, propagation, and pathloss modeling,''
  \emph{IEEE Wireless Commun. Lett.}, vol.~9, no.~5, pp. 581--585, 2019.

\bibitem{smart4}
C.~Huang, A.~Zappone, G.~C. Alexandropoulos, M.~Debbah, and C.~Yuen,
  ``Reconfigurable intelligent surfaces for energy efficiency in wireless
  communication,'' \emph{IEEE Trans. Wireless Commun.}, vol.~18, no.~8, pp.
  4157--4170, 2019.

\bibitem{smart5}
Q.~Wu and R.~Zhang, ``Intelligent reflecting surface enhanced wireless network
  via joint active and passive beamforming,'' \emph{IEEE Trans. Wireless
  Commun.}, vol.~18, no.~11, pp. 5394--5409, 2019.

\bibitem{smart6}
M.~{Di Renzo}, A.~Zappone, M.~Debbah, M.-S. Alouini, C.~Yuen, J.~D. Rosny, and
  S.~Tretyakov, ``Smart radio environments empowered by reconfigurable
  intelligent surfaces: How it works, state of research, and road ahead,''
  \emph{IEEE Journal on Selected Areas in Communications}, vol.~38, no.~11, pp.
  2450--2525, 2020.

\bibitem{Marco_proc}
M.~{Di Renzo}, F.~H. Danufane, and S.~Tretyakov, ``Communication models for
  reconfigurable intelligent surfaces: From surface electromagnetics to
  wireless networks optimization,'' \emph{https://arxiv.org/abs/2110.00833},
  2021.

\bibitem{Alu_2016}
N.~Mohammadi~Estakhri and A.~Al\`u, ``Wave-front transformation with gradient
  metasurfaces,'' \emph{Phys. Rev. X}, vol.~6, p. 041008, Oct 2016.

\bibitem{Elefth_2016}
A.~Epstein and G.~V. Eleftheriades, ``Synthesis of passive lossless
  metasurfaces using auxiliary fields for reflectionless beam splitting and
  perfect reflection,'' \emph{Phys. Rev. Lett.}, vol. 117, p. 256103, Dec 2016.

\bibitem{asadchy2016perfect}
V.~S. Asadchy, M.~Albooyeh, S.~N. Tcvetkova, A.~D{\'\i}az-Rubio, Y.~Ra'di, and
  S.~Tretyakov, ``Perfect control of reflection and refraction using spatially
  dispersive metasurfaces,'' \emph{Physical Review B}, vol.~94, no.~7, p.
  075142, 2016.

\bibitem{Diaz_From_2017}
A.~D\'{i}az-Rubio, V.~Asadchy, A.~Elsakka, and S.~Tretyakov, ``From the
  generalized reflection law to the realization of perfect anomalous
  reflectors,'' \emph{Science Advances}, vol.~3, no.~8, p. e1602714, 2017.

\bibitem{asadchy2017eliminating}
V.~S. Asadchy, A.~Wickberg, A.~D{\'i}az-Rubio, and M.~Wegener, ``Eliminating
  scattering loss in anomalously reflecting optical metasurfaces,'' \emph{ACS
  Photonics}, vol.~4, no.~5, pp. 1264--1270, 2017.

\bibitem{RA}
J.~Huang and J.~A. Encinar, \emph{Reflectarray Antennas}.\hskip 1em plus 0.5em
  minus 0.4em\relax Wiley, 2008.

\bibitem{Radi_Metagrating_2018}
Y.~Ra’di, D.~L. Sounas, and A.~Al{\`u}, ``Metagratings: Beyond the limits of
  graded metasurfaces for wave front control,'' \emph{Physical Review Letters},
  vol. 119, no.~6, 2018.

\bibitem{Do-Hoon_2018}
D.-H. Kwon, ``Lossless scalar metasurfaces for anomalous reflection based on
  efficient surface field optimization,'' \emph{IEEE Antennas and Wireless
  Propagation Letters}, vol.~17, no.~7, pp. 1149--1152, 2018.

\bibitem{Diaz_Power_2019}
A.~D{\'i}az-Rubio, J.~Li, C.~Shen, S.~Cummer, and S.~Tretyakov, ``Power flow
  conformal metamirrors for engineering wave reflections,'' \emph{Science
  Advances}, vol.~5, no.~2, p. eaau7288, 2019.

\bibitem{Xuchen_multiple}
X.~Wang, A.~D\'{\i}az-Rubio, and S.~A. Tretyakov, ``Independent control of
  multiple channels in metasurface devices,'' \emph{Phys. Rev. Applied},
  vol.~14, p. 024089, Aug 2020.

\bibitem{budhu2021perfectly}
J.~Budhu and A.~Grbic, ``Perfectly reflecting metasurface reflectarrays: Mutual
  coupling modeling between unique elements through homogenization,''
  \emph{{IEEE} Trans. Antennas Propag.}, vol.~69, no.~1, pp. 122--134, Jan.
  2021.

\bibitem{mailloux2017}
R.~J. Mailloux, \emph{Phased Array Antenna Handbook}, 3rd~ed.\hskip 1em plus
  0.5em minus 0.4em\relax Norwood, MA: Artech House, 2017.

\bibitem{Popov_Controlling_2018}
V.~Popov, F.~Boust, and S.~N. Burokur, ``Controlling diffraction patterns with
  metagratings,'' \emph{Physical Review Applied}, vol.~10, no.~1, p. 011002,
  2018.

\bibitem{Epstein_Unveiling_2017}
A.~Epstein and O.~Rabinovich, ``Unveiling the properties of metagratings via a
  detailed analytical model for synthesis and analysis,'' \emph{Physical Review
  Applied}, vol.~8, no.~5, p. 054037, 2017.

\bibitem{rabinovich_ieeejap2018}
O.~Rabinovich and A.~Epstein, ``Analytical design of printed circuit board
  (pcb) metagratings for perfect anomalous reflection,'' \emph{{IEEE} Trans.
  Antennas Propag.}, vol.~66, no.~8, pp. 4086--4095, Aug. 2018.

\bibitem{rabinovich_prb2019}
O.~Rabinovich, I.~Kaplon, J.~Reis, and A.~Epstein, ``Experimental demonstration
  and in-depth investigation of analytically designed anomalous reflection
  metagratings,'' \emph{Phys. Rev. B}, vol.~99, Mar. 2019, {Art. no.} 125101.

\bibitem{Hansen_WM}
T.~B. Hansen, ``Array of line sources that produces preselected plane waves,''
  \emph{Wave Motion}, vol. 106, p. 102791, 2021.

\bibitem{Estakhri_Wavefront_2016}
N.~Estakhri and A.~Al{\`u}, ``Wave-front transformation with gradient
  metasurfaces,'' \emph{Physical Review X}, vol.~6, no.~4, p. 041008, 2016.

\bibitem{DoHoon_Lossless_2018}
D.-H. Kwon, ``Lossless scalar metasurfaces for anomalous reflection based on
  efficient surface field optimization,'' \emph{{IEEE} Antennas Wireless
  Propag. Lett.}, vol.~17, no.~7, 2018.

\bibitem{Do-Hoon_planar_2021}
------, ``Planar metasurface design for wide-angle refraction using interface
  field optimization,'' \emph{{IEEE} Antennas Wireless Propag. Lett.}, vol.~20,
  no.~4, pp. 428--432, 2021.

\bibitem{yepes_apl2021}
C.~Yepes, M.~Faenzi, S.~Maci, and E.~Martini, ``Perfect non-specular reflection
  with polarization control by using a locally passive metasurface sheet on a
  grounded dielectric slab,'' \emph{Appl. Phys. Lett.}, vol. 118, p. 231601,
  2021.

\bibitem{rabinovich_ieeejap2020}
O.~Rabinovich and A.~Epstein, ``Arbitrary diffraction engineering with
  multilayered multielement metagratings,'' \emph{{IEEE} Trans. Antennas
  Propag.}, vol.~68, no.~3, pp. 1553--1568, Mar. 2020.

\bibitem{wong_prx2018}
A.~M.~H. Wong and G.~V. Eleftheriades, ``Perfect anomaloua reflection with a
  bipartite {Huygens'} metasurface,'' \emph{Phys. Rev. X}, vol.~8, Feb. 2018,
  {Art. no.} 011036.

\bibitem{Tereshin}
O.~Tereshin, V.~Sedov, and A.~Chaplin, \emph{Synthesis of antennas on
  wave-delay structures (in Russian)}.\hskip 1em plus 0.5em minus 0.4em\relax
  Moscow: Svyaz, 1980.

\bibitem{El_arxiv}
G.~Xu, S.~V. Hum, and G.~V. Eleftheriades, ``Extreme beam-forming with
  metagrating-assisted planar antennas,''
  \emph{https://arxiv.org/abs/2110.13000}, 2021.

\end{thebibliography}

\end{document}